\documentclass[aps,pra,twocolumn,showpacs]{revtex4}
\usepackage{amsmath,amssymb,graphicx,bbold}

\providecommand{\bra}[1]{\langle #1 \rvert}
\providecommand{\ket}[1]{\lvert #1 \rangle}

\begin{document}

\title{Fractional topological phase on spatially encoded photonic qudits}
\author{A. Z. Khoury$^1$, L. E. Oxman$^1$, B. Marques$^2$, 
A. Matoso$^2$, and S. P\'adua$^2$}

\affiliation{
1-Instituto de F\'\i sica, Universidade Federal Fluminense,
24210-346 Niter\'oi - RJ, Brasil\\
2-Departamento de F\'\i sica, Universidade Federal de Minas Gerais,
31270-901 Belo Horizonte - MG, Brasil.}
\date{\today}

\begin{abstract}
We discuss the appearance of fractional topological phases on cyclic 
evolutions of entangled qudits encoded on photonic degrees of freedom. 
We show how the spatial correlations between photons generated by 
spontaneous parametric down conversion can be used to evidence the 
multiple topological phases acquired by entangled qudits and the role 
played by the Hilbert space dimension. A realistic experimental proposal 
is presented with numerical predictions of the expected results. 
\end{abstract}

\maketitle

\section{introduction}

Geometrical phases are robust means for implementing unitary gates useful 
for quantum information processing \cite{vedral,zoller}. 
The phase evolution of entangled qubits was investigated 
in refs.\cite{sjoqvist,sjoqvist2}, and the 
role of entanglement on the topological nature of geometric phases 
has been discussed both theoretical \cite{remy,remy2,milman} and 
experimentally \cite{topoluff,nmr} in the context of two-qubit systems. 
More recently, it has been shown that the dimension of the Hilbert space 
plays a crucial role on the topological phases acquired by entangled qudits 
\cite{fracuff}. The appearance of fractional phases is a remarkable property 
of two-qudit systems, also shared by multiple qubits \cite{multiqubit}. However, 
experimental demonstrations of these interesting features are still rare. 
Fractional phases were originally investigated in quantum Hall systems in 
connection with different homotopy classes in the configuration space of anyons. 
Moreover, the topological origin of fractional phases in quantum Hall systems 
open interesting perspectives related to fault tolerant quantum computation 
\cite{faulttolerant}. 

In the present work we propose an experimental setup to observe fractional 
topological phases with entangled qudits encoded on the transverse position 
of quantum correlated photon pairs 
generated by spontaneous parametric down 
conversion (SPDC) \cite{sebastiao,sebastiao2,sebastiao3}. Local unitary operations 
can be applied in this degree of freedom with the aid of spatial light modulators 
that introduce a programmable phase profile on the wavefront of the quantum correlated 
photons. Then, the desired fractional phases can be observed through polarization 
controlled two-photon interference \cite{posinterf,imagepol}. The roles played by 
entanglement and Hilbert space dimension are unraveled by the behavior of 
the interference fringes as the unitary operations are continuously changed, raising 
the conjecture that entanglement and dimensionality witnesses are involved in 
the quantities to be measured. These ideas are supported by analytical and 
numerical calculations of the expected interference fringes. 
The paper is organized as follows: in section II we introduce the fractional 
topological phase concept; the experimental proposal is presented in section III; 
section IV is devoted to numerical simulations, and in section V conclusions are 
outlined.

\section{Fractional topological phases}

Let $\ket{\psi}=\sum_{i,j=1}^{d} \alpha_{ij}\ket{ij}$ be the most 
general two-qudit pure state. We shall represent this state by the 
$d\times d$ matrix $\mathbf{\alpha}$ whose elements are the coefficients 
$\alpha_{ij}$. With this notation the norm of the state vector becomes 
$\langle\psi|\psi\rangle=\mathrm{Tr}(\mathbf{\alpha^{\dagger}\alpha})=1$ and 
the scalar product between two states is 
$\langle\phi|\psi\rangle=\mathrm{Tr}(\mathbf{\beta^{\dagger}\alpha})$, where 
$\mathbf{\beta}$ is the $d\times d$ matrix containing the coefficients 
of state $\ket{\phi}$ in a given basis. 
In order to characterize a general vector in the Hilbert space, 
ote that any invertible matrix admits a polar decomposition
$\mathbf{\alpha}=e^{i\phi}\,Q\,S$, where $Q$ is a positive definite 
Hermitian matrix, and $S \in SU(d)$. Following the usual 
terminology of spontaneous parametric down conversion, we shall label 
the two qudits as \textit{signal} ($s$) and \textit{idler} ($i$). 
Under local unitary operations $U_s$ and $V_i$ the coefficient 
matrix is transformed according to
$
\mathbf{\alpha^{\prime}}=U_s\,\alpha\,V_i^T\;.
$
It can be readily seen that this kind of transformation preserves 
the polar decomposition and can be represented separately in 
each sector of the coefficient matrix: 
$
\mathbf{\alpha^{\prime}}=e^{i\phi^{\prime}}\,Q^{\prime}\,S^{\prime}\,
$,
where 
$e^{i\phi^{\prime}}=e^{i\phi}\sqrt[d]{\det{U_s}\cdot\det{V_i}\,}\,$, 
$Q^{\prime}=\bar{U}_s\,Q\,\bar{U}_s^{\dagger}\,$, and
$S^{\prime}=\bar{U}_s\,S\,\bar{V}_i^T\,$. $\bar{U}_s$ and 
$\bar{V}_i$ are the $SU(d)$ parts of the 
local unitary operations: $U_s=\sqrt[d]{\det{U_s}\,}\,\,\bar{U}_s$ and 
$V_i=\sqrt[d]{\det{V_i}\,}\,\,\bar{V}_i\,$.
Therefore, we identify the transformation in three  
sectors of the matrix structure: an explicit phase transformation 
$\phi\mapsto\phi^{\prime}$, a transformation closed in the space of positive definite 
Hermitian matrices $Q\mapsto Q^{\prime}$, and a transformation $S\mapsto S^{\prime}$ 
closed in $SU(d)$. Since the roots of a complex number are multivalued functions, 
some caution is necessary in defining the above quantities. For time varying 
unitary operations, we shall assume that any one of the possible roots is taken 
at the initial time, then the subsequent values must form a continuous evolution 
as a function of time, so that $\phi(t)$ is a smooth, everywhere differentiable 
function. 

Following \cite{mukunda}, we shall define as cyclic those evolutions for 
which the final state is physically equivalent to the initial one, in other 
words, the initial and final state vectors are related only by a global 
phase factor: $\mathbf{\alpha^{\prime}}=e^{i\theta}\mathbf{\alpha}\,$. 
The geometric phase acquired by a time evolving quantum state 
$\mathbf{\alpha}(t)=e^{i\phi(t)}\,Q(t)\,S(t)\,$
is always defined as
\begin{eqnarray}
\phi_g &=& \arg{\langle\psi(0)|\psi(t)\rangle} + 
i\int dt \,\,\langle\psi(t)|\dot{\psi}(t)\rangle 
\\
&=& \arg\left\{\mathrm{Tr}\left[\mathbf{\alpha^{\dagger}}(0)\mathbf{\alpha}(t)\right]\right\} 
+ i\int dt\,\, \mathrm{Tr}\left[\mathbf{\alpha^{\dagger}}(t)\dot{\mathbf{\alpha}}(t)\right]
\nonumber\;,
\label{phig}
\end{eqnarray}
that corresponds to the total phase minus the dynamical phase. At this point we 
can profit from the polar decomposition by investigating the contribution coming 
from each sector of the coefficient matrix. Let us consider a cyclic evolution 
over the time interval $T\,$: $\mathbf{\alpha}(T)=e^{i\theta}\mathbf{\alpha}(0)\,$. 
First, we identify the trivial phase evolution $\phi(T)=\phi(0)+\Delta\phi\,$. 
In the positive Hermitian sector, if we write $Q(T)=e^{i\theta_q}Q(0)$, its 
defining condition imposes $\theta_q=0\,$, so that no phase contribution could stem 
from this sector. Finally, in the $SU(d)$ sector we can have $S(T)=e^{i\theta_s}S(0)\,$, 
however, $\det{S(T)}=e^{id\theta_s}\det{S(0)}$, and since both $S(T)$ and $S(0)$ are 
$SU(d)$ matrices, we arrive at
\begin{equation}
\theta_s=\frac{2n\pi}{d}\;,
\label{thetas}
\end{equation}
where $n=0,1,2,...,d-1\,$. Therefore, only fractional phase values can arise from the 
$SU(d)$ sector and the global phase acquired in a cyclic evolution is 
\begin{equation}
\theta=\Delta\phi+\frac{2n\pi}{d}\;.
\label{theta}
\end{equation}
Although $SU(d)$ is simply connected, the topological nature of these fractional phases 
relies on the multiply connected manifold that represents the set of $SU(d)$ matrices 
$S$ with the identification $e^{i\theta_s}S\equiv S\,$. 

We can also investigate the dynamical phase using the 
polar decomposition. At this point it will be useful to write:
\begin{equation}
\dot{\mathbf{\alpha}}=i\dot{\phi}\,\mathbf{\alpha}(t)+e^{i\phi(t)}\left[\dot{Q}(t)S(t)+Q(t)\dot{S}(t)\right]\;.
\label{alphadot}
\end{equation}
This expression can be used in the dynamical phase integral appearing in eq.(\ref{phig}). Using the 
normalization condition $\mathrm{Tr}[\mathbf{\alpha^{\dagger}}(t)\mathbf{\alpha}(t)]=1\,$, we see 
that the first term in eq.(\ref{alphadot}) will cancel out the trivial global phase $\Delta\phi\,$, 
which do not contribute to the geometric phase, as would be expected. Then, by making 
$\mathbf{\alpha^{\dagger}}(t)=e^{-i\phi(t)}S^{\dagger}(t)Q(t)$ and using the cyclic property 
of the trace, one easily arrives at 
\begin{eqnarray}
\phi_g &=& \frac{2n\pi}{d} + i\,\int_0^T dt\,\, \mathrm{Tr}\left[Q(t)\dot{Q}(t)\right]
\nonumber\\
&+& i\,\int_0^T dt\,\, \mathrm{Tr}\left[Q^2(t)\dot{S}(t)S^{\dagger}(t)\right]\;.
\label{phig2}
\end{eqnarray}
The normalization condition implies in $\mathrm{Tr}[Q^2(t)]=1$. Differentiating both sides 
with respect to $t$ and using the cyclic property of the trace gives 
$\mathrm{Tr}\left[Q(t)\dot{Q}(t)\right]=0$, so that 
the first integral in eq.(\ref{phig2}) vanishes.
Now, it is now important to identify the following invariants under local unitary evolutions: 
$\mathrm{Tr}[\rho_j^{\, p}]$, $p=1,\dots, d$, where $\rho_j$ is the reduced density matrix 
of qudit $j\,$: 
\begin{equation}
\rho_s =(\alpha^{\dagger}\alpha)^T=(S^{\dagger}\,Q^2\,S)^T\;,
\label{rhos}
\end{equation}
and
\begin{equation}
\rho_i=\alpha \alpha^\dagger=Q^2\;.
\label{rhoi}
\end{equation}
In fact, the invariants are $j$-independent since one easily shows that 
$\mathrm{Tr}[\rho_s^{\, p}]=\mathrm{Tr}[\rho_i^{\, p}]=\mathrm{Tr}[Q^{2p}]\,$. The first one ($p=1$) is simply the 
norm of the state vector, as already stated. The second invariant is related to the 
{\it I-concurrence} of a two-qudit pure quantum state \cite{Iconc} 
$C=\sqrt{2(1-\mathrm{Tr}[\rho_j^2])}$, so that its invariance expresses the well 
known fact that entanglement is not affected by local unitary operations. 
The $p=d$ invariant can be rewritten in terms of the former and 
${\cal D}=|\det{\alpha}|$. In particular, for qubits we have 
$C=2\, {\cal D}$. For a given dimension $d$, the I-concurrence runs 
between $0$ for product states and $C_m=\sqrt{2(d-1)/d}$ for maximally entangled 
states.  
In order to exploit the role played by these invariants in the geometrical 
phase, we shall make them explicit in the expression of $Q^2$ in terms of 
the identity and the generators $T_n$ ($n=1,2,...,d^2-1$) of $SU(d)\,$. 
As usual, the generators are traceless matrices normalized according to 
$\mathrm{Tr}\,[T_nT_m]=\delta_{nm}/2\,$. A general $Q^2$ matrix with $\mathrm{Tr}[Q^2]=1$ 
and $\mathrm{Tr}[Q^4]=1-C^2/2$ can be written as
\begin{equation}
Q^2(t)= \frac{\mathbb{1}}{d} + \sqrt{C_m^2-C^2}\,\,\mathbf{\hat{q}}(t)\cdot \mathbf{T}\;,
\label{Q2}
\end{equation}
where $\mathbf{\hat{q}}$ is a unit vector in $\mathbb{R}^{d^2-1}$. 
For maximally entangled states ($C=C_m$), the reduced density matrices 
become $\rho_s=\rho_i=\mathbb{1}/d\,$. 

The expression (\ref{Q2}) can be used in the last term of eq. (\ref{phig2}) to establish a 
connection between the geometrical phase and the concurrence of the two-qudit state undergoing 
a cyclic evolution. However, it will be useful to work out a more convenient parametrization 
of $\dot{S}\,S^{\dagger}\,$. First we recall that for a general invertible matrix $\mathbb{A}$ 
we have \cite{cookbook}:
\begin{equation}
\frac{d\left(\det\mathbb{A}\right)}{dt}=\det\mathbb{A}\,\mathrm{Tr}\left[\mathbb{A^{-1}}\frac{d\mathbb{A}}{dt}\right]\;.
\end{equation}
Since the evolution of $S(t)$ is closed in $SU(d)$ ($\det S(t)=1$), we easily arrive at 
$\mathrm{Tr}\left[\dot{S}\,S^{\dagger}\right]=0\,$. Besides, note that $\dot{S}(t)S^{\dagger}(t)=-S(t)\dot{S}^{\dagger}(t)\,$, 
what allows us to define the \textit{velocity} vector ${\rm\bf s}\in\mathbb{R}^{d^2-1}$ such that 
$\dot{S}\,S^{\dagger}=i\,{\rm\bf s}\cdot\mathbf{T}\,$.  
Then, using the normalization condition for the generators and the fact that they are traceless matrices, 
we arrive at a compact expression for the geometrical phase:
\begin{eqnarray}
\phi_g &=& \frac{2n\pi}{d} - 
\frac{1}{2}\,\sqrt{C_m^2 - C^2}\,\oint \mathbf{\hat{q}}\cdot{\rm\bf dx}\;,
\label{phig3}
\end{eqnarray}
where we defined ${\rm\bf dx}={\rm\bf s}\,dt\,$. 
This key result shows that the 
geometrical phase for a cyclic evolution is composed by a fractional contribution of 
topological nature and an integral contribution which depends on the history of the 
quantum state evolution, parametrized in the Hermitian and $SU(d)$ sectors of the polar 
decomposition. This integral contribution is weighted by entanglement and vanishes 
completely for maximally entangled states, for which only the fractional values are 
allowed. We next turn to a possible physical implementation of the fractional phases 
using entangled qudits encoded on spatial degrees of freedom of photon pairs generated 
by spontaneous parametric down conversion (SPDC).

\section{Polarization controlled two-photon interference}

\begin{figure}
\begin{center} 
\includegraphics[scale=.5]{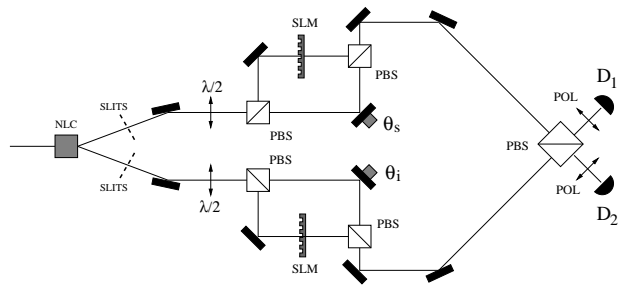}
\end{center} 
\caption{Experimental proposal. NLC: Nonlinear crystal, $\lambda/2$ half waveplate, 
PBS: Polarizing beam splitter, SLM: Spatial light modulator, POL: polarizer, 
$\theta_j$ ($j = i,s$) are path phases added by the mirrors coupled to PZTs and 
D$_l$ ($l = 1,2$) are single photon detectors.}
\label{setup}
\end{figure}

Let us consider the setup sketched in Fig.(\ref{setup}). Using multiple slit masks, 
spatial qudits are encoded on the transversal paths of the 
photon pairs generated by SPDC. 
Beyond the slits, the wave field can be written as a superposition 
of the field transmitted through different slits. We shall refer to these 
contributions as the \textit{slit modes} $\eta_m(\mathbf{r})$, where $m$ is the 
slit index. The slit mode functions satisfy the orthonormality condition: 
\begin{equation}
\int\,d^2\mathbf{r}\,\, \eta_m^*(\mathbf{r})\,\eta_n(\mathbf{r}) = \delta_{mn}\,. 
\end{equation}
The positive frequency components of signal and idler vector field operators are
\begin{eqnarray}
\mathbf{E}_s^{+}&=&E_{sH}^{+}\hat{e}_H + E_{sV}^{+}\hat{e}_V\;,
\nonumber\\
\mathbf{E}_i^{+}&=&E_{iH}^{+}\hat{e}_H + E_{iV}^{+}\hat{e}_V\;,
\end{eqnarray}
where $\hat{e}_{\mu}$ ($\mu=H,V$) is the horizontal ($H$) or vertical ($V$) polarization unit vector. 
Each polarization component is expanded in terms of the slit modes as
\begin{eqnarray}
E^+_{s\mu}&=&\sum_p a_{p\mu} \eta_p(\mathbf{r}_s)\;,
\nonumber\\
E^+_{i\nu}&=&\sum_q b_{q\nu} \eta_q(\mathbf{r}_i)\;,
\end{eqnarray}
where greek indices refer to polarization modes and roman ones refer to slit modes.
The annihilation operators $a_{p\mu}$ and $b_{q\nu}$ act on 
Fock states as usual:
\begin{eqnarray}
a_{p\mu}b_{q\nu}\ket{m\sigma,n\epsilon}&=&\delta_{pm}\delta_{\mu\sigma}\delta_{qn}\delta_{\nu\epsilon}\ket{vac}\,,
\label{aniq}
\end{eqnarray}
where $\ket{vac}$ is the vacuum state and $\ket{m\sigma,n\epsilon}$ is a 
two-photon Fock state corresponding to one 
signal photon passing through slit $m$ with polarization $\sigma$ and one 
idler photon passing through slit $n$ with polarization $\epsilon$. 

For type I phase matching and a vertically polarized pump beam, SPDC generates 
horizontally polarized photon pairs. Meanwhile, the spatial correlations 
between signal and idler can be tailored by playing with the 
pump angular spectrum \cite{sebastiao} using suitable lenses (omitted in the 
setup for simplicity) on the pump laser. These spatial correlations 
determine the two-photon path correlation through the slits. 
Therefore, the two-photon state after the slits can be written as
\begin{equation}
\ket{\psi_0}=\sum_{m,n} \alpha_{mn} \ket{mH,nH}\;,
\end{equation}
where $\alpha_{mn}$ is the probability amplitude of having one signal 
photon passing through slit $m$ and one idler photon passing through $n$. 
Right after the slits, two half wave plates ($\lambda /2$) rotate 
the photons polarization by $45^o$, and the two-photon quantum state 
becomes 
\begin{eqnarray}
\ket{\psi_1}&=&\sum_{m,n} \frac{\alpha_{mn}}{2}\left(\frac{}{} \ket{mH,nH}+\ket{mH,nV}\right.
\nonumber\\
&+&\left.\frac{}{}\ket{mV,nH}+\ket{mV,nV}\right)\;.
\end{eqnarray}
Then, 
each photon passes through a polarization controlled unitary gate composed 
by a Mach-Zehnder interferometer with input and output polarizing beam splitters (PBS), 
and a spatial phase modulator (SLM) inserted on the vertical polarization arm.  
The signal and idler evolutions $U_s$ and $V_i$ are then implemented on the vertical 
polarization component of the spatial qudits encoded on signal and idler photons, 
respectively. Of course, the efficient implementation of the unitary operations 
require that the slits are imaged on the SLM's. For the sake of simplicity, we 
shall omit the lenses required for this imaging scheme. 
After the controlled gates, the two-photon quantum state becomes
\begin{eqnarray}
\ket{\psi_2}&=&\sum_{m,n} \frac{\alpha_{mn}}{2}\left(\frac{}{}e^{i(\theta_s+\theta_i)}\,\ket{mH,nH}
+e^{i\theta_s}V_i\,\ket{mH,nV}\right.
\nonumber\\
&+&\left.\frac{}{}e^{i\theta_i}U_s\,\ket{mV,nH}+U_s\,V_i\,\ket{mV,nV}\right)\;,
\end{eqnarray}
where $\theta_s$ and $\theta_i$ are longitudinal phases added to signal and idler 
when they cross the Mach-Zehnder interferometers. The displacement of 
a piezoelectric ceramic (PZT) coupled to one of the Mach-Zenhder mirrors can do this task.

Finally, signal and idler are sent to different input ports of the same PBS so that 
only the components $\ket{nH,mH}$ and $\ket{nV,mV}$ will contribute to coincidences. 
Two polarizers (POL) oriented at $45^o$ are placed before detectors 
in order to \textit{erase} the polarization information. 
Then, the detected field operators are:
\begin{eqnarray}
E_1^+&=&\frac{1}{\sqrt{2}}\left(i E_{sV}^+ + E_{iH}^+\right)
\nonumber\\
E_2^+&=&\frac{1}{\sqrt{2}}\left( E_{sH}^+ + i E_{iV}^+\right)\;,
\label{e1e2}
\end{eqnarray}
and the coincidence count is proportional to:
\begin{eqnarray}
\bra{\psi_2} E^-_1E^-_2E^+_2E^+_1\ket{\psi_2}=\left\| E^+_2E^+_1\ket{\psi_2}\right\|^2\;, 
\end{eqnarray}
where $E_{j}^{-} = (E_{j}^{+})^{\dagger}$ and $j = 1,2\,$.
From eq.(\ref{aniq}), one easily sees that the $\ket{mH,nV}$ and $\ket{mV,nH}$ 
contributions vanish and the normalized coincidence function reduces to
\begin{eqnarray}
C(\mathbf{r}_1\,,\,\mathbf{r}_2)&=& 
\frac{1}{2}
\left\|\sum_{m,n}
e^{i\theta}\,\alpha_{mn}\,E^+_{iH} E^+_{sH}\,\ket{mH,nH}\right.
\nonumber\\
&+& \left.
\alpha_{mn}^{\prime}\,E^+_{iV} E^+_{sV}\,\ket{mV,nV}\frac{}{} \right\|^2
\;.
\end{eqnarray}
where $\theta=\theta_s + \theta_i - \pi\,$. The coefficients $\alpha^{\prime}_{mn}$ result 
from the local unitary transformations on the two-qudit state:  
\begin{equation}
U_s V_i\sum_{m,n} \alpha_{mn}\ket{mV,nV}=\sum_{m,n} \alpha^{\prime}_{mn} \ket{mV,nV}\; .
\end{equation}
The two-photon interference pattern is
\begin{eqnarray}
C(\mathbf{r}_1\,,\,\mathbf{r}_2)&=& 
\frac{1}{2}\left| \sum_{m,n} \eta_m(\mathbf{r}_1)\eta_n(\mathbf{r}_2)
\left(\alpha_{mn}e^{i\theta}+\alpha^{\prime}_{mn}\right)\right|^2
\,.
\nonumber\\
\end{eqnarray}

We are interested only in the phase acquired by the qudits when they evolve under the 
action of the unitary gates $U_s$ and $V_i$. 
Since the slit modes are orthonormal, the integrated coincidence count eliminates 
the spatial interference between different slits, 
so that only the Mach-Zehnder interference shows up in the two-photon correlations.
This corresponds to large aperture detection, insensitive to the detailed spatial 
structure of the two-photon quantum correlations. 
However, the Hilbert space dimension remains manifested through the 
two-qudit coefficients in the integrated coincidence function:
\begin{eqnarray}
C&\equiv& \int\,d^2\mathbf{r_1}\,\,d^2\mathbf{r_2}\,C(\mathbf{r}_1\,,\,\mathbf{r}_2) 
\nonumber\\
&=&
\frac{1}{2} \sum_{m,n} \left|\alpha_{mn}e^{i\theta}+\alpha^{\prime}_{mn}\right|^2\;.
\label{Calpha}
\end{eqnarray}

Note that polarization has been used as a subsidiary degree of freedom to provide 
two paths for the evolution of the spatial qudits. In fact, the interference 
described by eq.(\ref{Calpha}) can be thought of as resulting from the 
superposition between the initial two-qudit state $\ket{\varphi_0}=\sum_{mn}\alpha_{mn}\ket{mn}$ 
and the evolved state $\ket{\varphi}=U_sV_i\ket{\varphi_0}$ such that
\begin{eqnarray}
C&=&\left\|\frac{\ket{\varphi_0}+\ket{\varphi}}{\sqrt{2}}\right\|^2
\nonumber\\
&=&
\frac{1 + \left| \bra{\varphi_0}\varphi\rangle\right| \cos\left(\theta-\gamma\right)}{2}\;,
\label{Cvarphi}
\end{eqnarray}
where $\gamma=\arg\bra{\varphi_0}\varphi\rangle$. The visibility of the interference 
fringes is the absolute value of the state overlap 
$\bra{\varphi_0}\varphi\rangle\,$, while the interference phase is the 
overlap argument. For a cyclic evolution, $\ket{\varphi}= e^{i\gamma}\ket{\varphi_0}$, 
the interference recovers maximal visibility with the fringes shifted by $\gamma\,$. 
Finally, the topological phases can be implemented by continuously varying the 
unitary gates according to local time dependent parameters, 
adjusted by the signal and idler SLM's respectively. Next, we give 
some numerical examples of the possible operations and the expected interference 
patterns that can reveal both the fractional topological phases and the role 
of entanglement.

\section{Numerical Results}

We can illustrate the fractional phases with the simplest example of two qutrits 
($d=3$) under the action of diagonal unitary operations 
$U_s={\rm diag}[e^{i\phi_1},e^{i\phi_2},e^{i\phi_3}]$ and 
$V_i={\rm diag}[e^{i\chi_1},e^{i\chi_2},e^{i\chi_3}]\,$, 
where $\phi_n$ and $\chi_n$ are phase factors imposed 
by the spatial light modulators (SLM) to the photon path states. 
In order to guarantee that no trivial dynamical phase is added 
by the operations, we assume that $\sum_n\phi_n=\sum_n\chi_n=0\,$, 
in other words, we restrict the two-qutrit evolutions to 
$SU(3)\otimes SU(3)$ operations. 

The coefficient matrix evolves as 
$\mathbf{\alpha^{\prime}}=U_s\,\mathbf{\alpha}\,V_i^{T}$, so that 
\begin{eqnarray}
\alpha^{\prime}_{mn}&=&e^{i[\phi_m+\chi_n]}\,\,\alpha_{mn}\;,
\end{eqnarray}
and the coincidence count becomes
\begin{eqnarray}
C&=&\frac{1}{4}\sum_{m,n}\,\left|\alpha_{mn}\right|^2\,
\left|e^{i\theta}+e^{i[\phi_m+\chi_n]}\right|^2
\nonumber\\
&=&\sum_{m,n}\,\left|\alpha_{mn}\right|^2\,\cos^2\left[\frac{\phi_m+\chi_n-\theta}{2}\right]\;.
\label{Cphi}
\end{eqnarray}

In principle, the two SLMs can be operated independently and implement totally different 
local evolutions. However, for simplicity, we shall first consider an identical operation capable 
to evidence both the fractional phases and the signatures of entanglement. Our cyclic 
operations will be decribed by a real parameter $t$ in the interval $t\in [0,1]\,$.
As a first example, let us consider
\begin{eqnarray}
\phi_1=\chi_1&=&\frac{\pi}{3} \left[2t-\left(2t-1\right)\,H\left(t-\frac{1}{2}\right)\right]\;,
\nonumber\\
\phi_2=\chi_2&=&-\frac{2\pi}{3}\,t\;,
\nonumber\\
\phi_3=\chi_3&=&\frac{\pi}{3}\left(2t-1\right)\,H\left(t-\frac{1}{2}\right)\;.
\label{phiH}
\end{eqnarray}
where $H(t)$ is the Heaviside function. Eqs.(\ref{phiH}) decribe a continuous phase evolution in 
each qudit component, corresponding to a local $SU(d=3)$ operation independently applied to each qudit. 
Under this $SU(3)\otimes SU(3)$ operation, we may compare a maximally entangled state:
\begin{eqnarray}
\ket{\varphi_e}=\frac{1}{\sqrt{3}}\left(\ket{00}+\ket{11}+\ket{22}\right)\,,
\label{phient}
\end{eqnarray}
($\alpha_{mn}=\delta_{mn}/\sqrt{3}$) with a product state having the same single qudit population distribution:
\begin{eqnarray}
\ket{\varphi_p}=\frac{1}{3}\left(\ket{0}+\ket{1}+\ket{2}\right)\otimes\left(\ket{0}+\ket{1}+\ket{2}\right)\,,
\label{phiprod}
\end{eqnarray}
($\alpha_{mn}=1/3$). 
%
\begin{figure}
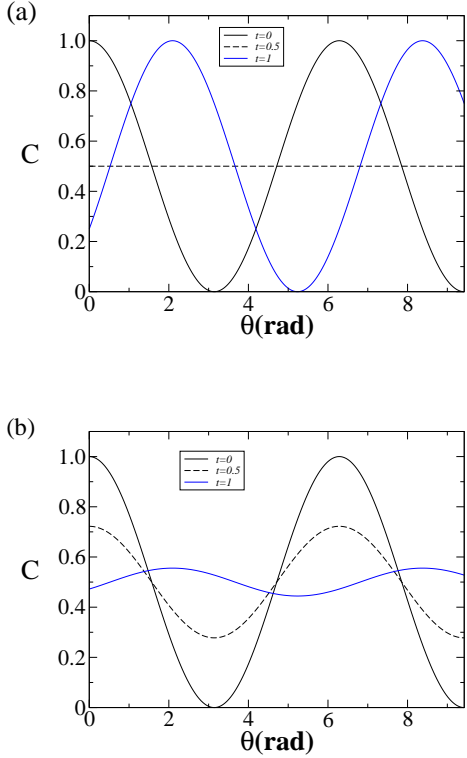

\begin{center} 
\includegraphics[scale=.25]{fig2a.eps}\\
\vspace{1cm}
\includegraphics[scale=.25]{fig2b.eps}
\end{center} 
\caption{Coincidence interference for (a) a maximally entangled state and 
(b) a product state, corresponding to the unitary operations given by eqs.(\ref{phiH}), 
for three values of $t$: $t = 0$ (continuous line), $t = 0.5$ (dashed line) and $t = 1$ (blue line).}
\label{graph1}
\end{figure}
%
The resulting coincidence interference given by eq.(\ref{Cphi}) is shown in 
Figs. \ref{graph1}a and \ref{graph1}b. As we can see, the 
behaviour is considerably different for product and entangled states. For the entangled state, the interference 
fringes completely disappear when $t=1/2$, and the two-qudit entangled state becomes orthogonal to the 
initial state. Then, the fringes reappear with the fractional phase shift $2\pi/3$, reaching maximal visibility again 
at $t=1$. On the other hand, the product state exhibits a remarkable different behaviour, the fringes never disappear, 
and maximal visibility is not recovered when $t=1\,$. 

\begin{figure}
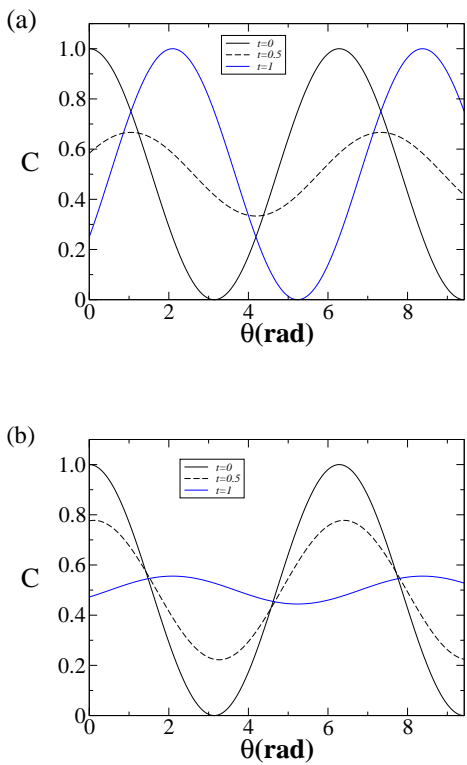

\begin{center} 
\includegraphics[scale=.25]{fig3a.eps}\\
\vspace{1cm}
\includegraphics[scale=.25]{fig3b.eps}
\end{center} 
\caption{Coincidence interference for (a) a maximally entangled state and 
(b) a product state, corresponding to the unitary operations given by eqs.(\ref{phicont}), 
for three values of $t$: $t = 0$ (continuous line), $t = 0.5$ (dashed line) and $t = 1$ (blue line).}
\label{graph2}
\end{figure}

Let us now consider a different kind of evolution controlled by independent parameters $t_s$ and $t_i$:
\begin{eqnarray}
\phi_1&=&\phi_2=\frac{\pi}{3}\,t_s\;,
\nonumber\\
\phi_3&=&-\frac{2\pi}{3}\,t_s\;,
\nonumber\\
\chi_1&=&\chi_2=\frac{\pi}{3}\,t_i\;,
\nonumber\\
\chi_3&=&-\frac{2\pi}{3}\,t_i\;.
\label{phicont}
\end{eqnarray}
In this case, the coincidence is
\begin{eqnarray}
C_e=\frac{2}{3}\cos^2\left(\frac{t\pi}{3}-\frac{\theta}{2}\right)+
\frac{1}{3}\cos^2\left(\frac{2t\pi}{3}+\frac{\theta}{2}\right)\;,
\label{Ce}
\end{eqnarray}
for the entangled state of eq.(\ref{phient}), and 
\begin{eqnarray}
C_p&=&\frac{4}{9}\cos^2\left(\frac{t\pi}{3}-\frac{\theta}{2}\right)+\frac{1}{9}\cos^2\left(\frac{2t\pi}{3}+\frac{\theta}{2}\right)
\nonumber\\
&+&\frac{2}{9}\left[1+\cos\left(\pi\tau\right)\,\,\cos\left(\frac{t\pi}{3}+\theta\right)\right]\;,
\label{Cp}
\end{eqnarray}
for the product state of eq.(\ref{phiprod}), where we defined $t=(t_s+t_i)/2$ and $\tau=(t_s-t_i)/2\,$. 
As we can see, the coincidence interference is completely different in each case. 
It does not depend on the relative parameter $\tau$ for the entangled input, 
while it does for the product one. The behaviour of the fringe visibility 
is also considerably different, even for $t_s=t_i=t$ ($\tau=0)\,$. In Fig. \ref{graph2} the coincidence interference 
is plotted for $\tau=0$ and different values of $t\,$. In this case, the evolved entangled state never becomes 
orthogonal to the initial one, but the fringes recover maximum visibility for $t=1$, with the 
fractional phase shift $2\pi/3$. However, the same operation does not take the product state through a 
cyclic evolution and the associated interference never recovers maximum visibility. 

\begin{figure}[hb]
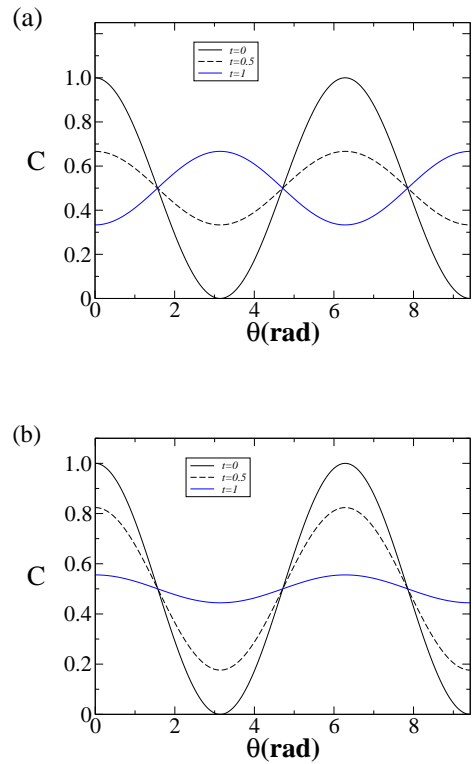

\begin{center} 
\includegraphics[scale=.25]{fig4a.eps}\\
\vspace{1cm}
\includegraphics[scale=.25]{fig4b.eps}
\end{center} 
\caption{Coincidence interference for (a) a maximally entangled state and 
(b) a product state, corresponding to the unitary operations given by eqs.(\ref{phiH2}), 
for three values of $t$: $t = 0$ (continuous line), $t = 0.5$ (dashed line) and $t = 1$ (blue line).}
\label{graph3}
\end{figure}

Finally, in order to observe the effect of the Hilbert space dimension, it is instructive 
to consider a situation in which only two components of the qutrits are operated, and compare 
their evolution with a genuine pair of qubits. For example, let us make:
\begin{eqnarray}
\phi_1&=&-\phi_2=\frac{\pi}{2}\,t_s\;,
\nonumber\\
\phi_3&=&0\;,
\nonumber\\
\chi_1&=&-\chi_2=\frac{\pi}{2}\,t_i\;,
\nonumber\\
\chi_3&=&0\;,
\label{phiH2}
\end{eqnarray}
In this case, the coincidence is
\begin{eqnarray}
C_e=\frac{1}{2}+\frac{\cos\theta}{6}\left[1+2\cos(\pi\,t)\right]\;,
\label{Ce2}
\end{eqnarray}
for the entangled state (\ref{phient}), and 
\begin{eqnarray}
C_p&=&\frac{1}{2}+\frac{\cos\theta}{9}\left[\frac{1}{2}+\cos\left(\pi\,t\right)
+\cos\left(\pi\,\tau\right)\right.
\nonumber\\
&+&2\left.\cos\left(\frac{\pi}{2}\,t\right)\cos\left(\frac{\pi}{2}\,\tau\right)\frac{}{}\right]\;,
\label{Cp2}
\end{eqnarray}
for the product state (\ref{phiprod}). As before, the coincidence count associated to an 
entangled input is independent of the relative control parameter $\tau$, while the same 
does not hold true for the product input. However, as can be readily seen in Figs. \ref{graph3}a 
and \ref{graph3}b, neither the entangled nor the product inputs are driven through a cyclic evolution in 
the full range of the total control parameter $t\,$. The interference fringes never 
recover maximal visibility. 

\begin{figure}
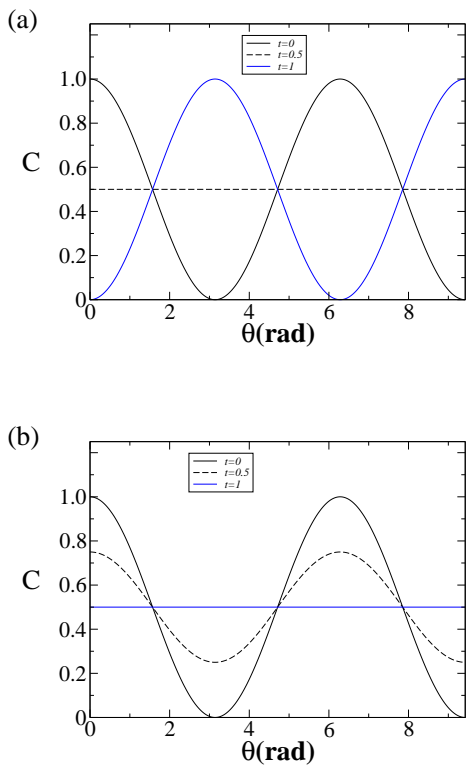

\begin{center} 
\includegraphics[scale=.25]{fig5a.eps}\\
\vspace{1cm}
\includegraphics[scale=.25]{fig5b.eps}
\end{center} 
\caption{Coincidence interference for a (a) maximally entangled state and 
(b) product state, corresponding to the unitary operations given by eqs.(\ref{phiHd2}).
$t = 0$ (continuous line), $t = 0.5$ (dashed line) and $t = 1$ (blue line).}
\label{graph4}
\end{figure}

We can compare these results with those expected for a pair of qubits ($d=2$) prepared 
either in the entangled state
\begin{eqnarray}
\ket{\varphi_e}=\frac{1}{\sqrt{2}}\left(\ket{00}+\ket{11}\right)\;,
\label{phientd2}
\end{eqnarray}
($\alpha_{mn}=\delta_{mn}/\sqrt{2}$) or in the product state having the same single qubit population distribution:
\begin{eqnarray}
\ket{\varphi_p}=\frac{1}{2}\left(\ket{0}+\ket{1}\right)\otimes\left(\ket{0}+\ket{1}\right)\;,
\label{phiprodd2}
\end{eqnarray}
($\alpha_{mn}=1/2$), evolving under 
\begin{eqnarray}
\phi_1&=&-\phi_2=\frac{\pi}{2}\,t_s\;,
\nonumber\\
\chi_1&=&-\chi_2=\frac{\pi}{2}\,t_i\;.
\label{phiHd2}
\end{eqnarray}
In this case, the coincidence is
\begin{eqnarray}
C_e(d=2)&=&\frac{1}{2}\left[1+\cos\left(\pi\,t\right)\cos\theta\right]\;,
\label{Ced2}
\end{eqnarray}
for the entangled state (\ref{phientd2}), and 
\begin{eqnarray}
C_p(d=2)&=&\frac{1}{2}+\frac{\cos\theta}{4}\left[\cos\left(\pi\,t\right) + 
\cos\left(\pi\,\tau\right)\right]\;,
\label{Cpd2}
\end{eqnarray}
for the product state (\ref{phiprodd2}). These coincidence counts are shown 
in Figs. \ref{graph4}a and \ref{graph4}b for $\tau=0\,$ and the full range of $t\,$. Now, the 
entangled state is taken through a cyclic evolution and recovers maximal visibility 
with the topological phase $\pi$, as expected for maximally entangled qubits 
\cite{remy,remy2,milman,topoluff,nmr,fracuff}. On the other hand, the product state 
does not complete a cyclic evolution, it ends up in a state orthogonal to the 
initial one. Therefore, 
the interference behaviour is not only sensitive to entanglement but also to 
the dimension of the qudits being operated. This leads to a natural conjecture 
that the quantities being measured are related to entanglement and dimensionality 
witnesses, although we shall leave the formal investigation of this aspect to 
a future work. 

\section{conclusions}

As is well-known, qudit gates based on topological phases are a potentially 
robust means for implementing quantum algorithms \cite{bullock,ashok,munro}. 
In this work, 
we proposed a physical implementation of fractional topological phases acquired 
by entangled qudits operated by local unitary transformations.
Spatial qudits encoded on quantum correlated photon pairs are a suitable framework 
for this investigation, since the qudit components may be efficiently addressed 
by spatial light modulators. The expected coincidence 
interferences can reveal both, the role played by entanglement as well as the Hilbert 
space dimension. Then, a natural conjecture arises as to whether the measured 
quantities are related to entanglement and dimensionality witnesses, what 
motivates further investigation.  

\section*{Acknowledgments}
We are grateful to E. Sj\"oqvist and M. Johansson for useful 
discussions. Funding was provided by Instituto Nacional de 
Ci\^encia e Tecnologia de Informa\c c\~ao Qu\^antica (INCT-CNPq), 
Coordena\c c\~{a}o de Aperfei\c coamento de Pessoal de N\'\i vel 
Superior (CAPES), Funda\c c\~{a}o de Amparo \`{a} Pesquisa do 
Estado do Rio de Janeiro (FAPERJ-BR), and Funda\c c\~{a}o de Amparo 
\`{a} Pesquisa do Estado de Minas Gerais (FAPEMIG-BR).

\end{document}